\documentclass[10pt,conference]{IEEEtran}
\usepackage{cite}
\usepackage{amsmath,amssymb,amsfonts}
\usepackage{algorithmic}
\usepackage{graphicx}
\usepackage{textcomp}
\usepackage{xcolor}
\usepackage{tabularx}
\usepackage{booktabs}
\def\BibTeX{{\rm B\kern-.05em{\sc i\kern-.025em b}\kern-.08em
    T\kern-.1667em\lower.7ex\hbox{E}\kern-.125emX}}
\ifCLASSOPTIONcompsoc
    \usepackage[caption=false, font=normalsize, labelfont=sf, textfont=sf]{subfig}
\else
\usepackage[caption=false, font=footnotesize]{subfig}
\fi

\date{}

\begin{document}
\title{Conformal Metasurfaces for Recovering Dynamic Blockage in Vehicular Systems}
\author{\IEEEauthorblockN{Marouan Mizmizi\IEEEauthorrefmark{1}, Dario Tagliaferri\IEEEauthorrefmark{1}, Misagh Khosronejad\IEEEauthorrefmark{1}, \\ Laura Resteghini\IEEEauthorrefmark{2}, Gian Guido Gentili\IEEEauthorrefmark{1}, Lorenza Draghi\IEEEauthorrefmark{1}, and Umberto Spagnolini\IEEEauthorrefmark{1}}
\IEEEauthorblockA{\IEEEauthorrefmark{1}Politecnico di Milano, Milan, Italy}\IEEEauthorblockA{\IEEEauthorrefmark{2}Huawei Technologies Italia S.r.l., Segrate, Italy} 
E-mail (corresponding):\,dario.tagliaferri@polimi.it}

\maketitle

\begin{abstract}
Intelligent reflecting surfaces (IRS) will represent a key technology in the upcoming sixth-generation (6G) communication networks to extend the network coverage and overcome link blockage. Research on IRS is expected to take a giant leap in the coming years to address the current technological limitations, mainly regarding the IRS re-configuration in highly dynamic scenarios, such as vehicle-to-vehicle (V2V) ones. This paper proposes a fully passive and low-cost solution based on pre-configured IRS to be lodged on the vehicle's body, which does not require any signaling for re-configuration. However, conventional IRS are planar array and cannot fit with most vehicles' silhouettes. Hence, the proposed design targets conformal surfaces (C-IRS) with an arbitrary shape. In particular, this paper reveals the first experimental findings on the realization of C-IRS operating at 26 GHz and measurements in anechoic chamber, which validate the analytical derivations. To demonstrate the benefits of the proposed solution, numerical simulations in a V2V scenario show that, when the percentage of vehicles equipping a C-IRS increases, the communication becomes more robust to vehicles' blockage and the average end-to-end SNR is enhanced up to 25 dB.
\end{abstract}


\section{Introduction}
The 5G network rollouts are in full swing globally, with standardization advancing to address new market verticals~\cite{3GPPTR38901}.
While optimization of early 5G devices and networks is an ongoing process, the research community is already focusing on what 6G networks will be. 
6G networks are expected to use higher frequencies in the millimeter-wave (mmWave - $30-100$ GHz), sub-THz, and THz bands \cite{rappaport2019wireless}. However, because of the harsher radio wave propagation, i.e., higher path and penetration losses compared to currently deployed system at sub-6 GHz \cite{jameel2018propagation}, high frequency radio links are particularly vulnerable to blockage, especially when the mobility of users increases (e.g., in vehicle-to-everything (V2X) scenarios \cite{dong2021vehicular}), resulting in limited communication range and unreliable links. %
In vehicle-to-vehicle (V2V) scenarios, the main source of blockage is another vehicle interposing between the transmitting vehicle (Tx) and the receiving vehicle (Rx), which can cause a power loss of up to $20$ dB at $30$ GHz, degrading with Tx-Rx distance and traffic intensity~\cite{dong2021vehicular}. 
%

6G network optimization will rely upon relays and smart metasurfaces to mitigate the blockage. Relay-based solutions represent one possibility, either using relays of opportunity~\cite{linsalata2021map} or macro-diversity at infrastructure level~\cite{Panwar8643739}. However, the latter pushing for network densification increases the cost of deployment and complexity. Differently, smart metasurfaces, a.k.a. reconfigurable intelligent surfaces (RIS) or intelligent reflecting surfaces (IRS) \cite{direnzo2021surveyRIS}, are an emerging technology capable of manipulating an incoming electromagnetic (EM) field to control the reflection/refraction direction.
Incorporating RIS/IRS into wireless networks has been advocated as a revolutionary means to transform any passive wireless propagation environment into an active controllable one, i.e., enabling the smart radio environment (SRE) \cite{zhang2021performance}. 
Herein, we refer to RIS when the amplitude and phase of each element can be set in real-time \cite{abeywickrama2020intelligent} and to IRS when it is pre-configured and not tunable~\cite{oliveri2021holographic}. Although the RIS and IRS terminologies are often used interchangeably in literature, we herein stress this distinction to distinguish between fully passive surfaces (IRS) and nearly passive ones (RIS). Moreover, RIS are more flexible in dynamic environments but require dedicated control signaling, while IRS are cheaper and need no power supply and control signaling.
\begin{figure}[b!]
\vspace{-0,4cm}
    \centering
    \includegraphics[width=0.6\columnwidth]{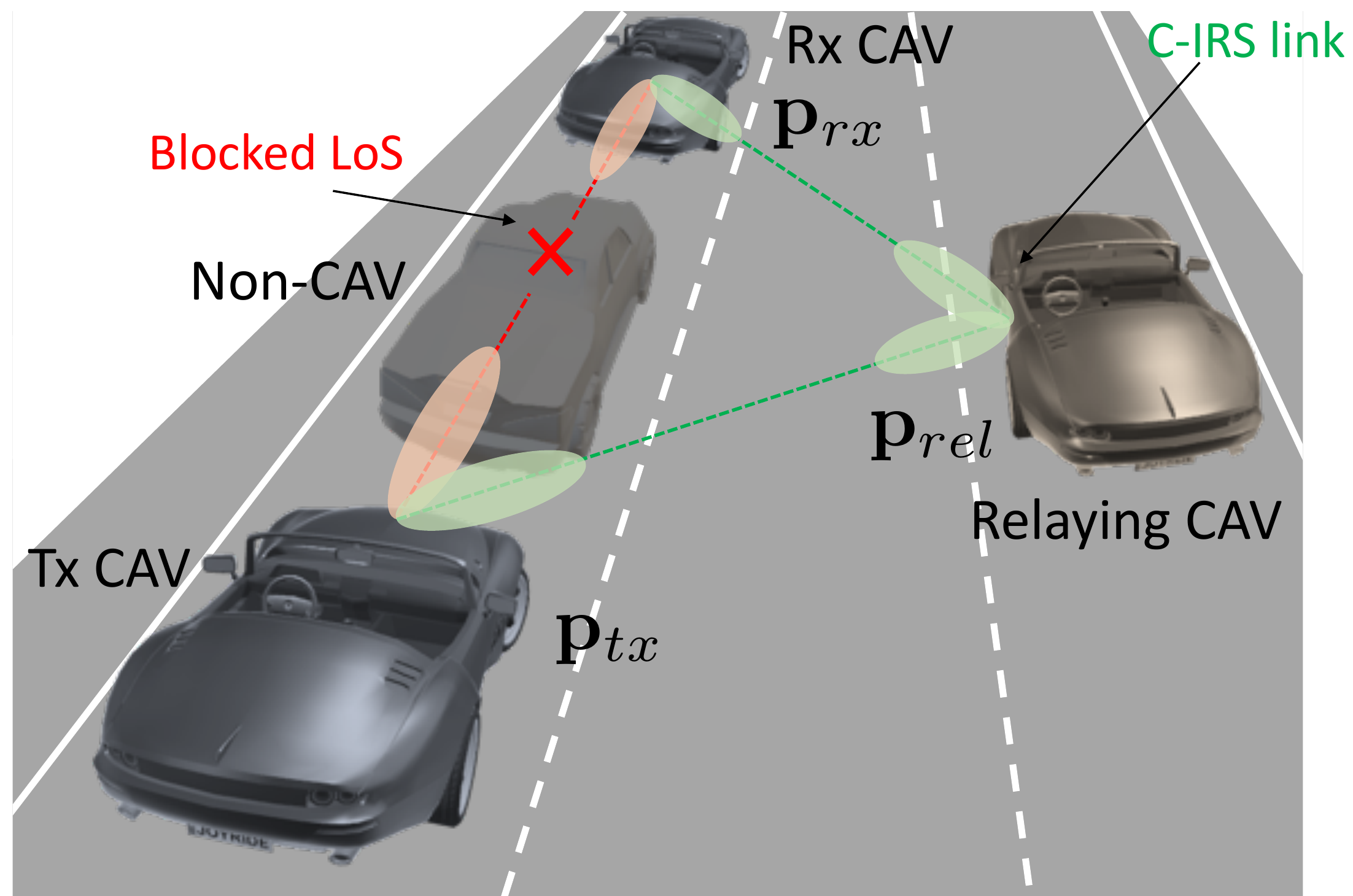}
    \caption{Highway scenario: CAVs equipped with C-IRS can serve as passive and low-cost relays to mitigate the blockage affection from non-CAV.}
    \label{fig:intro}
\end{figure}

RISs have been largely investigated in literature as a technology that improves the average capacity per unit area \cite{direnzo2021surveyRIS} and addresses the blockage issue in vehicular networks. Referring to blockage management, in \cite{Zeng9539048}, the authors propose a RIS-assisted handover scheme using deep reinforcement learning to mitigate the blockage in a cellular scenario. In the considered setting, the base station computes the RIS configuration for each cluster of users based on the observed channels. The proposed approach does not explicitly solve a non-convex optimization problem for RIS configuration as in \cite{chen2021qos}, yet, it still requires additional signaling overhead, which might be unsuitable for high mobility V2V scenario. Moreover, the real-time RIS reconfiguration is still an open issue, although some works consider a finite phase set at each RIS element to decrease the control overhead \cite{yin2020single}. 
Most works consider the perfect channel state information (CSI) available at the RIS, or they assume slowly varying CSI scenarios \cite{Zhou9110587}. However, for V2V links, the CSI outdates rapidly \cite{garcia2021tutorial}, making the state-of-the-art solutions inefficient or even impractical.
A novel hybrid RIS is considered in \cite{Idban-HRIS} to reduce the signaling overhead, where the RIS can sense the received power and self-configure. However, the hybrid solution requires additional hardware and power costs.
Conversely, IRS-based solutions provide a fully passive alternative to RIS when the deployment cost is prohibitive, or the incidence/reflection angles are not a-priori known and cannot be estimated accurately. 

In this regard, in our recent work \cite{mizmizi2022conformal}, we introduced a novel arrangement, which consists of mounting metasurfaces on next generation vehicles' sides, e.g., connected automated vehicles (CAVs), such that to mitigate the blockage affection by non-CAVs. However, since the vehicle shape is not planar, the standard RIS/IRS solution is not appropriate. Hence, we have generalized the configuration problem of RIS/IRS by considering a conformal design for any arbitrary 3D metasurface (C-RIS/C-IRS), and we derived the related analytical phase configuration. This paper extends previous works with the first experimental findings on the fabrication of a C-IRS (pre-configured) operating at 26 GHz and the corresponding radiation pattern measurements in anechoic chamber. The observed outcomes validate the analytical results (both lossless and lossy simulations) and thereby open up the opportunity for large-scale deployment of this technology since the production and maintenance cost is much more competitive than any other solution. As an additional result, we show the trend of the average SNR in a multi-lane highway scenario where a variable percentage of CAVs is equipped with C-RIS/C-IRS on their sides. Compared to conventional V2V systems exploiting only the direct link, C-IRS provide a beneficial SNR gain that grows with traffic density and CAVs percentage (up to $25$ dB), as well as C-RIS, that represent the benchmark and the upper performance bound (up to $40$ dB of SNR gain). These results further justify the adoption of conformal metasurfaces for 6G vehicular networks.

\textit{Organization}: The remainder of the paper is organized as follows: Section \ref{sect:system_model} outlines the system and channel model, and describes the proposed C-IRS design, Section \ref{sect:CIRS_realization} shows the practical realization and testing of the C-IRS while Section \ref{sect:V2X_results} validates the C-IRS benefits in a V2V network. Finally, conclusions are in Section \ref{sect:conclusion}.

\textit{Notation}: Bold upper- and lower-case letters describe matrices and column vectors. Matrix transposition and hermitian are indicated respectively as $\mathbf{A}^{\mathrm{T}}$ and $\mathbf{A}^{\mathrm{H}}$. $\mathrm{tr}\left(\mathbf{A}\right)$ extracts the trace of $\mathbf{A}$. $\mathrm{diag}(\mathbf{a})$ is the diagonal matrix given by vector $\mathbf{a}$. $\mathbf{I}_n$ is the identity matrix of size $n$. With  $\mathbf{a}\sim\mathcal{CN}(\boldsymbol{\mu},\mathbf{C})$ we denote a multi-variate circularly complex Gaussian random variable $\mathbf{a}$ with mean $\boldsymbol{\mu}$ and covariance $\mathbf{C}$. $\mathbb{E}[\cdot]$ is the expectation operator, while $\mathbb{R}$ and $\mathbb{C}$ stand for the set of real and complex numbers, respectively. $\delta_{n}$ is the Kronecker delta.

\section{System Model and C-IRS Design}\label{sect:system_model}

\begin{figure}[b!]
    \centering
    \includegraphics[width=\columnwidth]{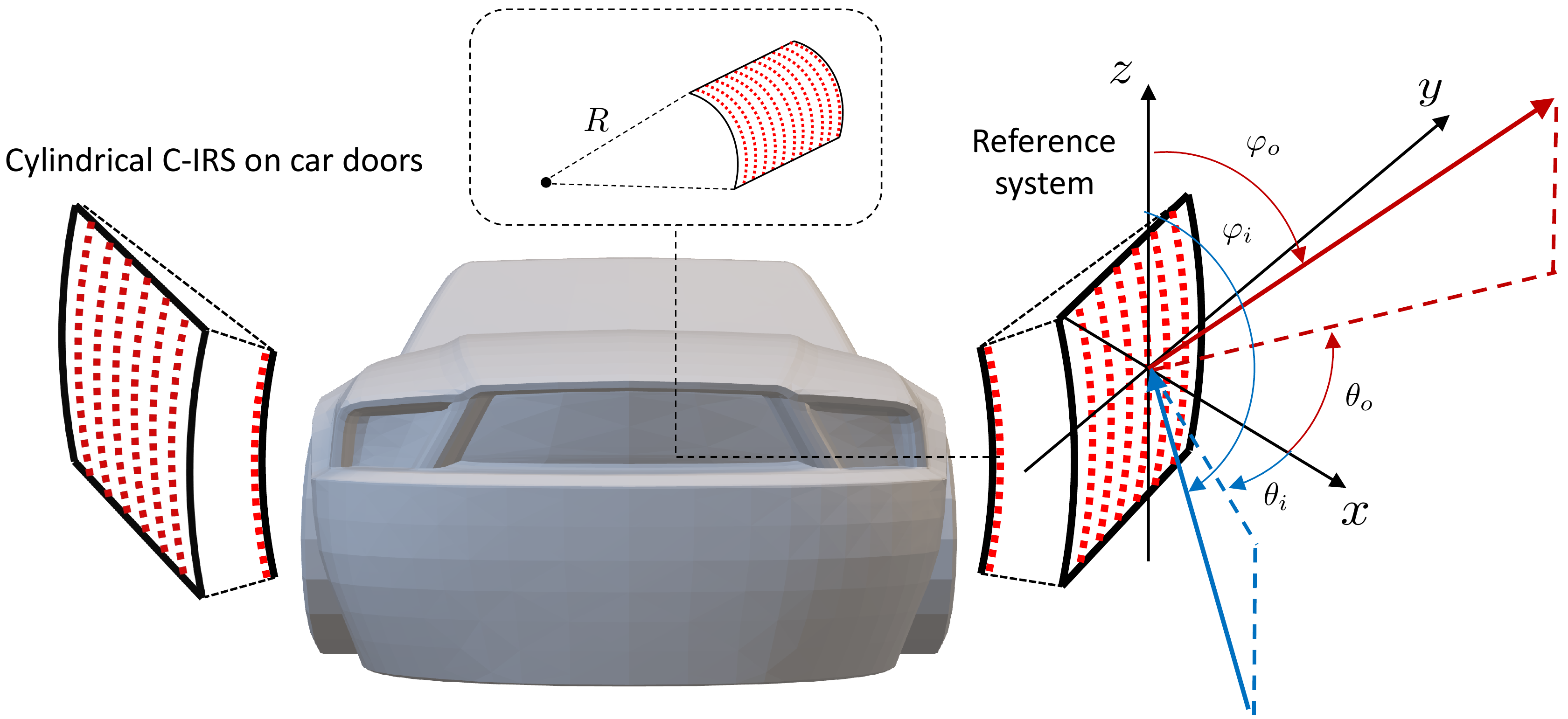}
    \caption{Sketch of the C-IRS reference system}
    \label{fig:systemModel}
\end{figure}

We consider the V2V network in Fig. \ref{fig:intro} representing a scenario with multiple lanes. A fraction $P$ of the vehicles are connected autonomous vehicles (CAVs), 
equipped with a mmWave/sub-THz uniform linear array (ULA) with $K$ antenna elements, and two C-IRS on both vehicle's sides. 
The position of Tx and Rx CAVs' antenna is $\mathbf{p}_{tx} = [x_{tx}, y_{tx}, z_{tx}]^\mathrm{T}$ and $\mathbf{p}_{rx} = [x_{rx}, y_{rx}, z_{rx}]^\mathrm{T}$, respectively, defined in the global coordinate system. Similarly, the candidate relaying CAV is located in $\mathbf{p}_{rel} = [x_{rel}, y_{rel}, z_{rel}]^\mathrm{T}$, where $\mathbf{p}_{rel}$ identifies the position of the C-IRS phase center. The global coordinate system is such that the cars move along direction $y$, the cross-motion axis is $x$, and $z$ denotes the vertical direction.

%
Herein, the shape of each C-IRS along the vehicles' sides is approximated by a cylinder section with a given radius $R$ and length $L$. Note that the cylindrical approximation allows to obtain close-form expressions of the phase pattern configuration, that generalize the results of planar IRS, which can be obtained for $R \rightarrow \infty$. Each C-IRS has $M$ elements deployed along the conformal coordinate (i.e., vehicle height $H$), and $N$ elements along the cylindrical direction (i.e., vehicle length $L$). The equipment is sketched in~Fig. \ref{fig:systemModel}. The position of the $(m,n)$-th C-IRS element, for $m = -M/2, \dots, M/2-1$ and $n = -N/2,\dots, N/2-1$, is $\mathbf{p}_{m,n} = \mathbf{p}_{rel} + [x_{m,n},\,y_{m,n},\,z_{m,n}]^\mathrm{T}$, the relative 3D displacement of the $(m,n)$-th element is: $x_{m,n} = R (\cos \psi_m-1)$, $y_{m,n} = d_n (n-1)$ and $z_{m,n} = R \sin\psi_m$. $\psi_m$ is the angular position in local cylindrical coordinates of the $m$-th row, while $d_m$ and $d_n$ are the elements' spacing along the vertical and horizontal directions, respectively. The C-IRS area is $A = L \times 2R \,\psi_{M} \approx L H$, where $\psi_{M} = M\arcsin\left(d_m/2/R\right)$ is the angular sector spanned by the C-IRS. According to the size of common car doors \cite{DoorRadius}, for $H=1$ m and $L=1$ m, and selecting $R$ in the range $[1,10]$ provides a reasonable closed-form shape approximation. 

\subsection{Signal Model}\label{subsect:signal_model}

Let the input-output relation of the MIMO communication system between Tx and Rx CAVs be
\begin{equation}\label{eq:receivedSignal}
    y = \mathbf{w}^\mathrm{H} \mathbf{H} \mathbf{f} \,s + \mathbf{w}^\mathrm{H} \mathbf{n},
\end{equation}
where $s$ is the complex Tx symbol of power $\sigma_s^2$, $\mathbf{n} \sim \mathcal{CN}\left(\mathbf{0},  \sigma^2_n\mathbf{I}_K\right)$ is the additive white Gaussian noise at the Rx antennas $\mathbf{f} \in \mathbb{C}^{K \times 1}$, $\mathbf{w} \in \mathbb{C}^{K \times 1}$ are the Tx and Rx beamforming vectors and 
\begin{equation}\label{eq:channelDefinition}
    \mathbf{H} = \mathbf{H}_{d} + \sum_{c=1}^{C} \mathbf{H}_{c,rx} \boldsymbol{\Phi} \mathbf{H}_{tx,c}
\end{equation}
is the spatially-sparse MIMO channel. The latter is the sum of the direct Tx-Rx channel $\mathbf{H}_{d}\in \mathbb{C}^{K \times K}$ and $C$ reflection channels via as many relaying CAVs equipped with C-IRS~\cite{Sanguinetti9300189}. In this setting, $\mathbf{H}_{tx,c}\in \mathbb{C}^{MN  \times K}$ is the channel from the Tx to the $c$-th C-IRS, $\mathbf{H}_{c,rx}\in \mathbb{C}^{K \times MN}$ the one from the $c$-th C-IRS and
\begin{equation}\label{eq:reflectingmatrix}
    \boldsymbol{\Phi} = \mathrm{diag}\left([ e^{j\Phi_{1}},...,e^{j\Phi_{mn}},..., e^{j\Phi_{MN}}]\right) \in \mathbb{C}^{MN \times MN}
\end{equation}
is the diagonal phase-configuration matrix at each C-IRS. With \eqref{eq:reflectingmatrix}, we assume for simplicity a lossless C-IRS made with a fully reflecting material with no mutual coupling between elements. The direct Tx-Rx channel is modeled under plane wave approximation as~\cite{rappaport2019wireless} 
\begin{equation}\label{eq:channelModel}
    \mathbf{H}_{d}  = \alpha_{d}\, \varrho_{rx}(\boldsymbol{\vartheta}_{d}^{rx}) \varrho_{tx}(\boldsymbol{\vartheta}_{d}^{tx}) \mathbf{a}_{rx}(\boldsymbol{\vartheta}_{d}^{rx})\mathbf{a}^\mathrm{H}_{tx}(\boldsymbol{\vartheta}_{d}^{tx}) 
\end{equation}
where \textit{(i)} $\alpha_d$ denotes the complex amplitude of the direct path, modelled as in \cite{3GPPTR37885}; \textit{(ii)} $\mathbf{a}_{tx}(\boldsymbol{\vartheta}_d^{tx})\in\mathbb{C}^{K\times 1}$, $\mathbf{a}_{rx}(\boldsymbol{\vartheta}_d^{rx})\in\mathbb{C}^{K\times 1}$ are the Tx and Rx array response vectors, function of the angles of arrival (AoAs) and departure (AoDs) of the direct path, respectively $\boldsymbol{\vartheta}_d^{rx} = (\theta_d^{rx}, \varphi_d^{rx})$ and $\boldsymbol{\vartheta}_d^{tx} = (\theta_d^{tx}, \varphi_d^{tx})$ (for azimuth and elevation), \textit{(iv)} $\varrho^{tx}(\boldsymbol{\vartheta}_d^{tx})$ and $\varrho^{rx}(\boldsymbol{\vartheta}_d^{rx})$ are the Tx and Rx single-antenna gains, respectively~\cite{9569465}. 

The model in \eqref{eq:channelModel} applies also to the reflection channels $\mathbf{H}_{tx,c}$ and $\mathbf{H}_{c,rx}$ whenever the Tx-relay and relay-Rx distance is in excess of few tens of meters. If this is the case, i.e. far-field condition, \eqref{eq:channelModel} holds with proper scattering amplitudes and \textit{local} AoAs and AoDs on/from the $c$-th C-IRS, routinely called angles of incidence (AoI) and reflection (AoR), denoted with $\boldsymbol{\vartheta}_{m,n}^{i,c}$ and $\boldsymbol{\vartheta}_{m,n}^{o,c}$ respectively. Local AoI/AoR can be expressed as function of the \textit{global} ones $\boldsymbol{\vartheta}^c_{i}=(\theta^c_i,\varphi^C_i)$ and $\boldsymbol{\vartheta}^c_{o}=(\theta^c_o,\varphi^c_o)$ knowing the shape of the C-IRS.

\subsection{Design of C-IRS for Vehicular Applications}\label{sect:CIRS_vehicular}

To lodge a C-IRS on the sides of any vehicle, it must match the shape of the vehicle's side silhouette, possibly thin (see Section \ref{sect:CIRS_realization}) and thus takes into account the surface's profile. 

Let us consider the C-IRS local reference system in Fig. \ref{fig:systemModel}. The goal is to design the C-IRS phase to reflect the impinging radio signal from direction $\boldsymbol{\vartheta}_i = (\theta_i,\varphi_i)$ towards $\boldsymbol{\vartheta}_o =(\theta_o,\varphi_o)$. Leveraging on the generalized Snell's law of reflection~\cite{PhysRevApplied.9.034021}, the C-IRS phase configuration across the elements is reported in \eqref{eq:phaseConfiguration}, for an arbitrarily shaped C-IRS.
\begin{figure*}[!t]
\begin{equation}\label{eq:phaseConfiguration}
    \Phi_{m,n} = - (2\pi/\lambda)\left[ x_{m,n}\, (\cos \theta_o \sin \varphi_o + \cos \theta_i \sin \varphi_i) + 
    y_{m,n} \,(\sin \theta_o \sin \varphi_o + \sin \theta_i \sin \varphi_i) +z_{m,n}\, (\cos \varphi_o + \cos \varphi_i)\right]
    \end{equation}
    \hrulefill
\end{figure*}
According to \eqref{eq:phaseConfiguration}, the C-IRS phase configuration requires the perfect knowledge of AoIs $\boldsymbol{\vartheta}_i$ and AoRs $\boldsymbol{\vartheta}_o$. However, they both quickly vary in a vehicular scenario and thus require a real-time reconfiguration, which is signaling intensive and might be unfeasible in practice. Hence, the proposed solution consists of realizing a C-IRS that behaves as an EM flat mirror, which can be obtained by setting a pre-configuration that matches the typical AoIs and AoRs in vehicular networks. As detailed in our recent work \cite{mizmizi2022conformal}, the elevation AoI/AoR ($\varphi_i$ and $\varphi_o$) are approximately Gaussian distributed around $\varphi=90$ deg with a typical standard deviation of $1.5$ deg, while the distribution of azimuth AoI/AoR ($\theta_i$ and $\theta_o$) is wider and monotonically increases toward $\theta \rightarrow \pm 90$ deg.

Therefore, we reasonably impose $\varphi_i=\varphi_o=90$ deg and leave the azimuth as a free design parameter $\theta_o=-\theta_i=\overline{\theta}$. By substituting the C-IRS elements positions in \eqref{eq:phaseConfiguration} we obtain:
\begin{equation}\label{eq:phasePerpendicular}
    \Phi_m(\overline{\theta}) = -\frac{4 \pi R}{\lambda} \left(\cos\psi_m -1 \right)\cos\overline{\theta},
\end{equation}
which depends \textit{only} on the shape of the C-IRS along the conformal dimension, i.e., on the curvature radius $R$ and angular position of the elements $\psi_m$, as well as on the operating azimuth angle $\overline{\theta}$ and wavelength $\lambda$. Notice that the phase gradient is non-zero along the curved dimension only, and that every other C-IRS shape can be accounted for by \eqref{eq:phaseConfiguration}.

\section{Realization of C-IRS}\label{sect:CIRS_realization}

\begin{figure}[b!]
	\centerline{\includegraphics[width=0.5\columnwidth]{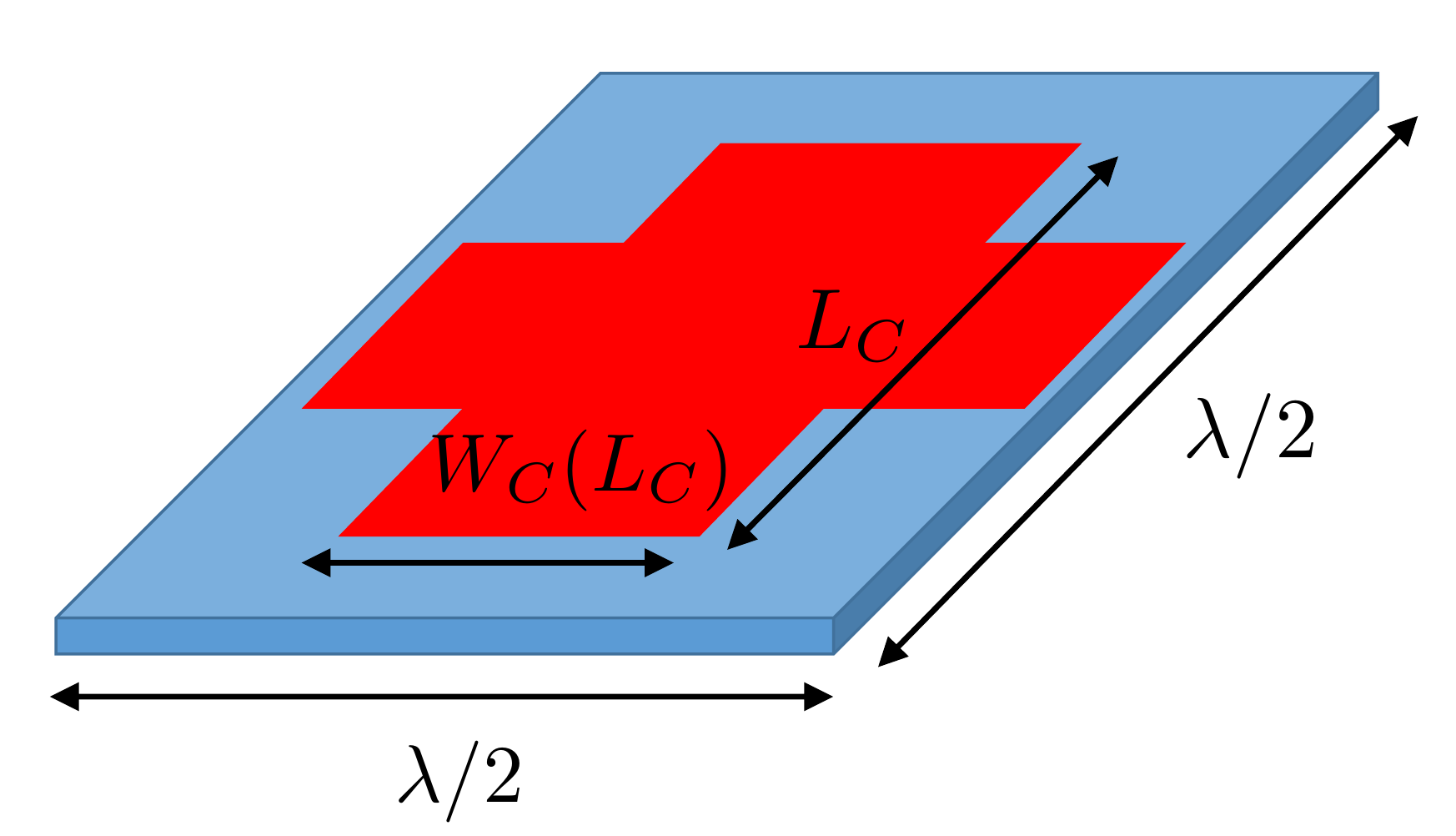}}
	\caption{The cross-shaped element used in the C-IRS.}\label{fig:G1}
\end{figure}
\begin{figure}[b!]
	\centerline{\includegraphics[width=0.9\columnwidth]{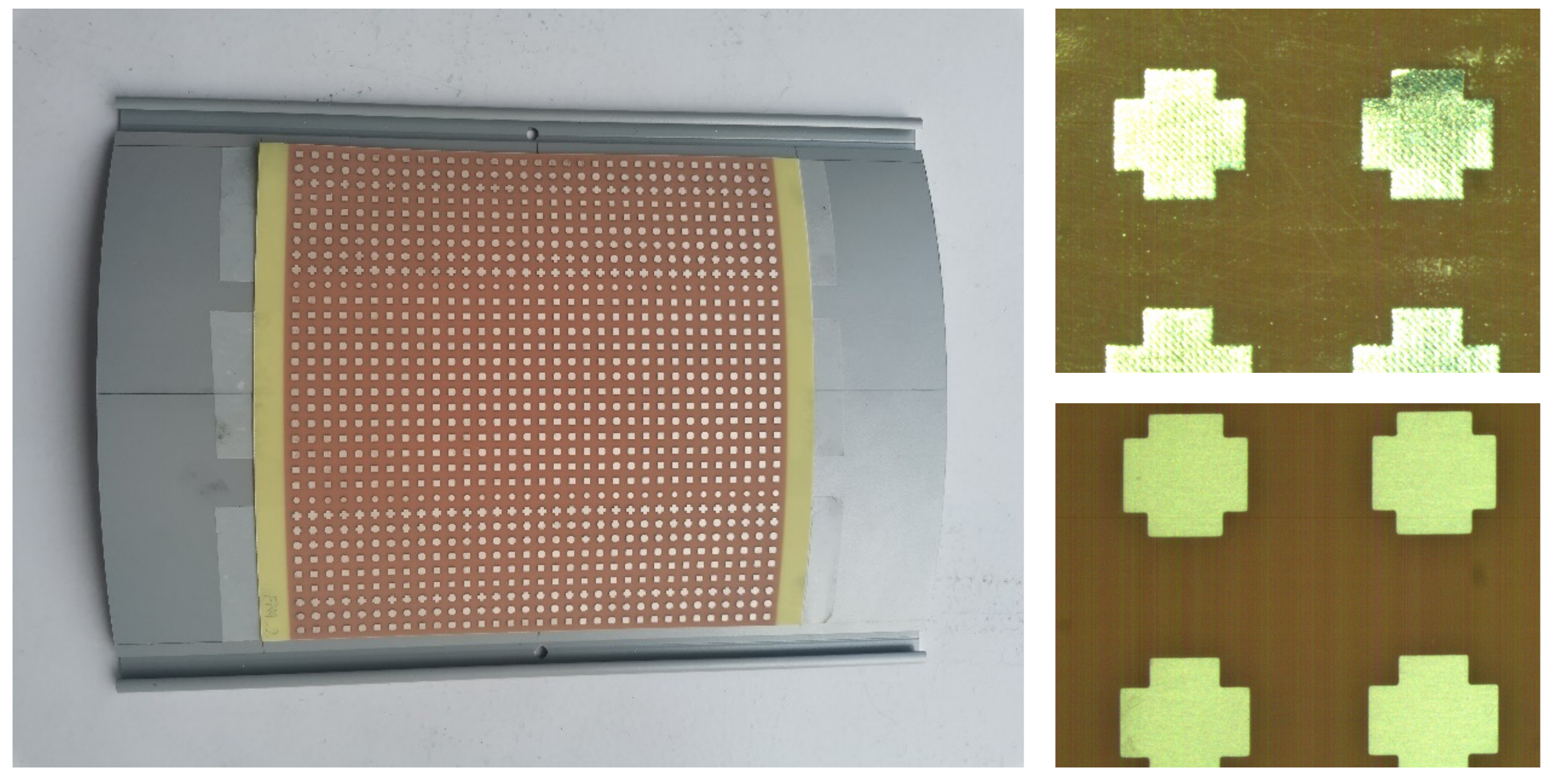}}
	\caption{The C-IRS and a closeup of two different realizations.}\label{fig:G1a}
\end{figure}
\begin{figure}[b!]
	\centerline{\includegraphics[width=0.95\columnwidth]{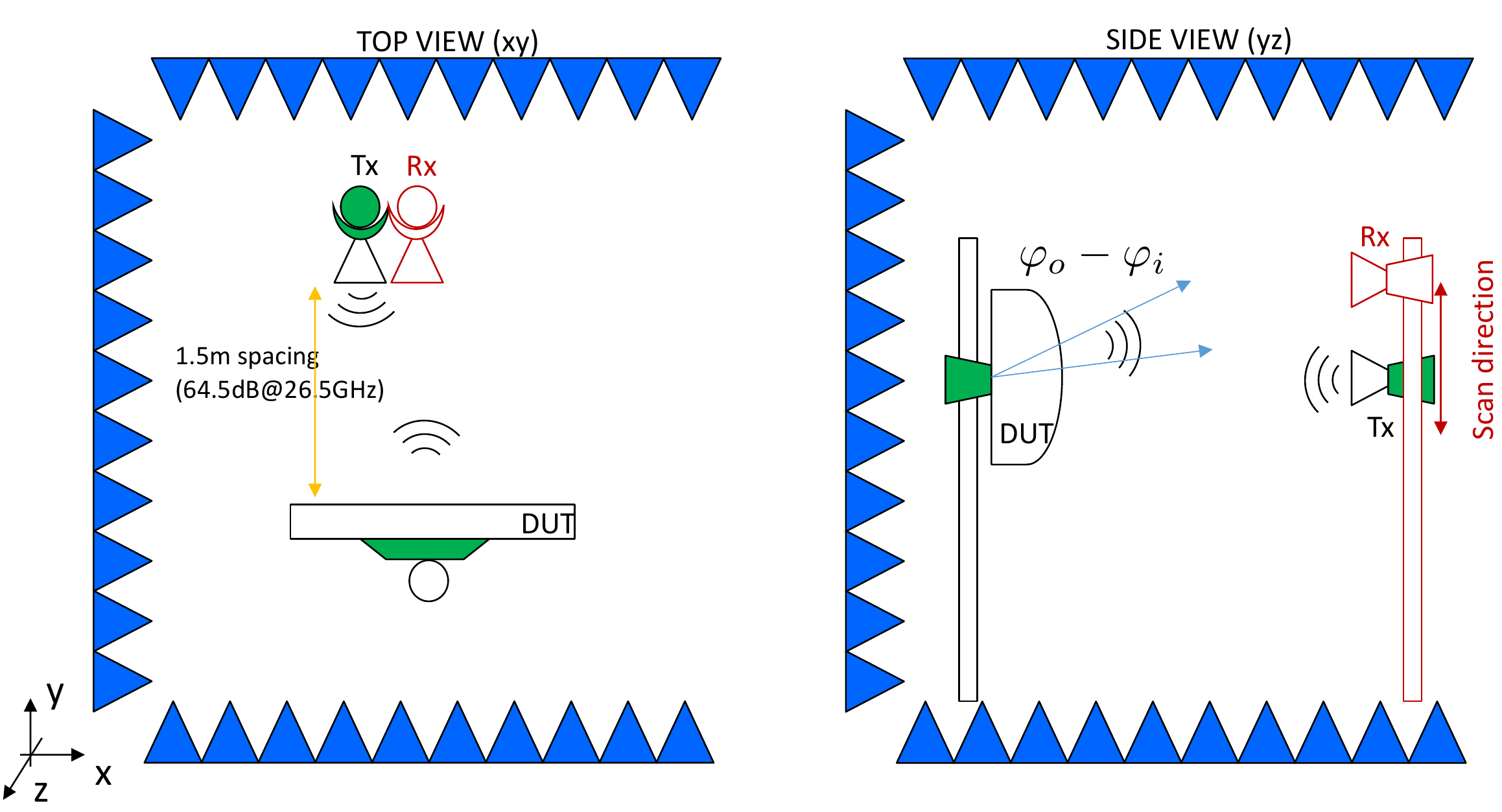}}
	\caption{The measurement setup for the scattered field.}\label{fig:G3a}
\end{figure}

C-IRS realization can be carried out in several ways, either using the concept of true metasurface~\cite{G1}, with an ideally continuum phase distribution, or using the concept of reflectarray~\cite{G2}, in which the phase distribution is sampled and each element of the reflectarray must introduce the phase required at the corresponding sampling point. The former approach provides a good degree of flexibility in the phase pattern design at the price of being complex while the latter one is by far the most used one for its simplicity and its better potential for massive and low cost production. The C-IRS design is practically ruled by the design of the master element. Reflectarray design discussed in this work is based on planar metallic resonating elements printed on a low cost flexible substrate. The design of the specific element must account for some potentially critical issues, such as tolerances in the realization, bandwidth, polarization (single or double) and ohmic efficiency. Optimal design is achieved by a combined effect of thickness of substrate material and shape of the planar element. Cross-shaped elements have been used here in the design (Fig. \ref{fig:G1}). These represent an excellent compromise between all requirements, because the parameters that control the cross ($W_C$ and $L_C$ in Fig. \ref{fig:G1}) can be optimized in terms of tolerances, bandwidth and phase value required by the selected element. In particular, it was found that by a suitable selection of these two parameters, the cross outperforms the simple square element in a double polarization system, both in terms of bandwidth and tolerances in fabrication. The final design uses different crosses with different values of $L_C$ and $W_C$ at different position, as a result of a global optimization of the required phases. The high resolution pattern was fabricated with a traditional photolithographic technique. Lower resolution pattern was prepared using a toner-reactive metal foil, hot laminated over a transfer paper printed in a standard laser printer.
We specifically accounted in the design for robustness to small lithographic errors in the realization. This requirement is achieved by designing the slope of the phase curve in the neighborhood of the resonant frequency. Excessive steepness in the phase curve introduces excessive sensitivity to small errors in the lithographic printing process and small differences of the permittivity of the substrate with respect to nominal value. Metal patterns have been printed by an inkjet printer. A quite good adhesion to the substrate was found even for very small thickness of the metallization. The ink adhere to the substrate and it forms tiny bubbles. Tolerances in the printing process do have an impact on the final design, but they did not severely impact the final performances.

As a manufacturing benchmark, we designed a C-IRS aiming to compensate a curved surface with a curvature radius of $30$ cm and working at $26$ GHz.
The C-IRS in the application  must provide a focused beam in the incidence plane where the curvature lies (see Fig. 2), thereby compensating the spreading of the wavefield due to the curvature of the surface. This specific design represents a good benchmark because here the radius of curvature is rather small for vehicles' door and the required element phases spans the complete set of angles from $-180$ to $180$ deg. The C-IRS occupies an area of $20 \times 20 $ cm$^2$ and $999$ elements are present on the curved surface. A view of the C-IRS is shown in Fig. \ref{fig:G1a}, together with a closeup of part of the patterned surface in two realizations.

\begin{figure}[!b]
	\centerline{\includegraphics[width=0.4\columnwidth]{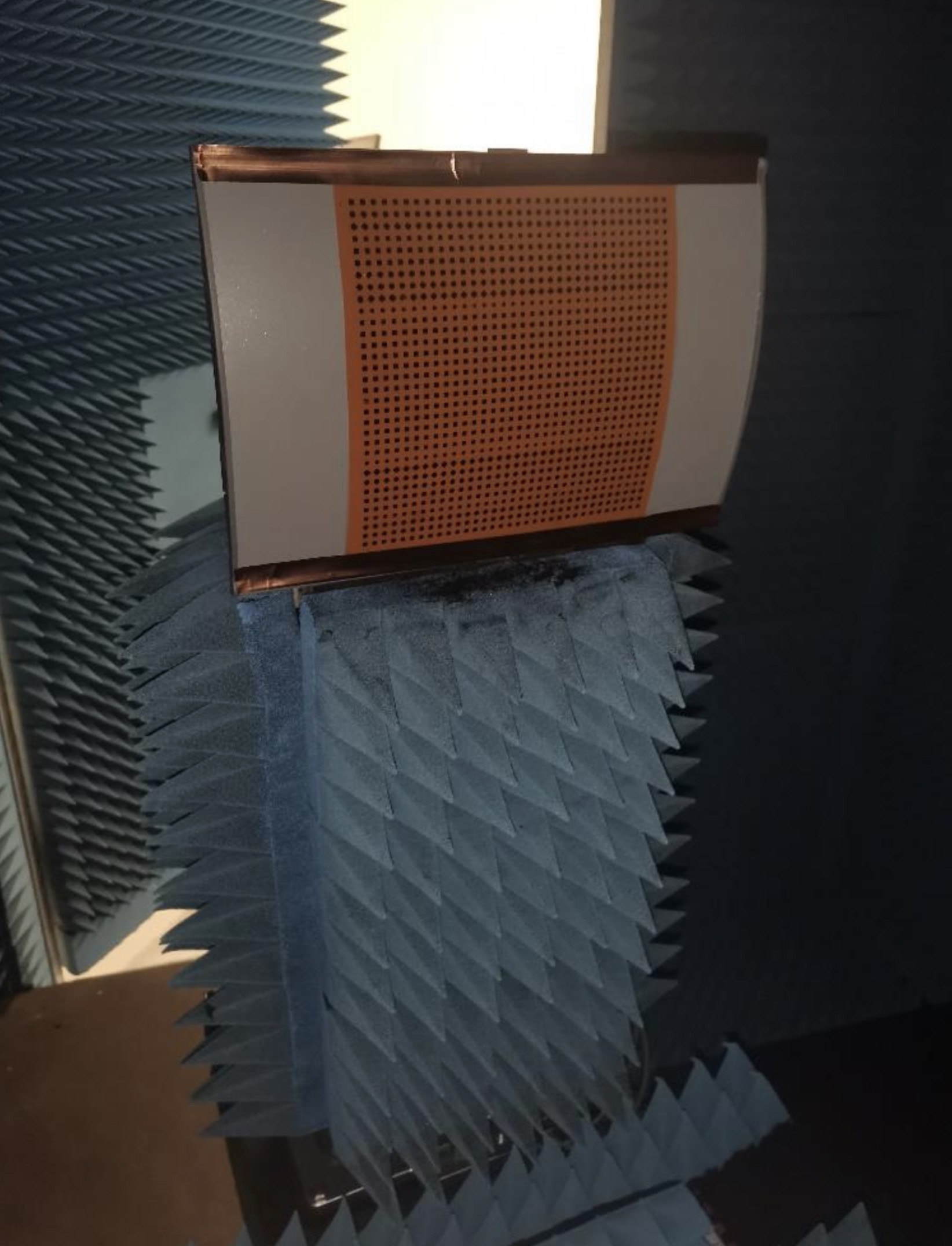}}
	\caption{A detail of the C-IRS mounted in the chamber.}\label{fig:G3}
\end{figure}

\begin{figure}[!b]
	\centerline{\includegraphics[width=\columnwidth]{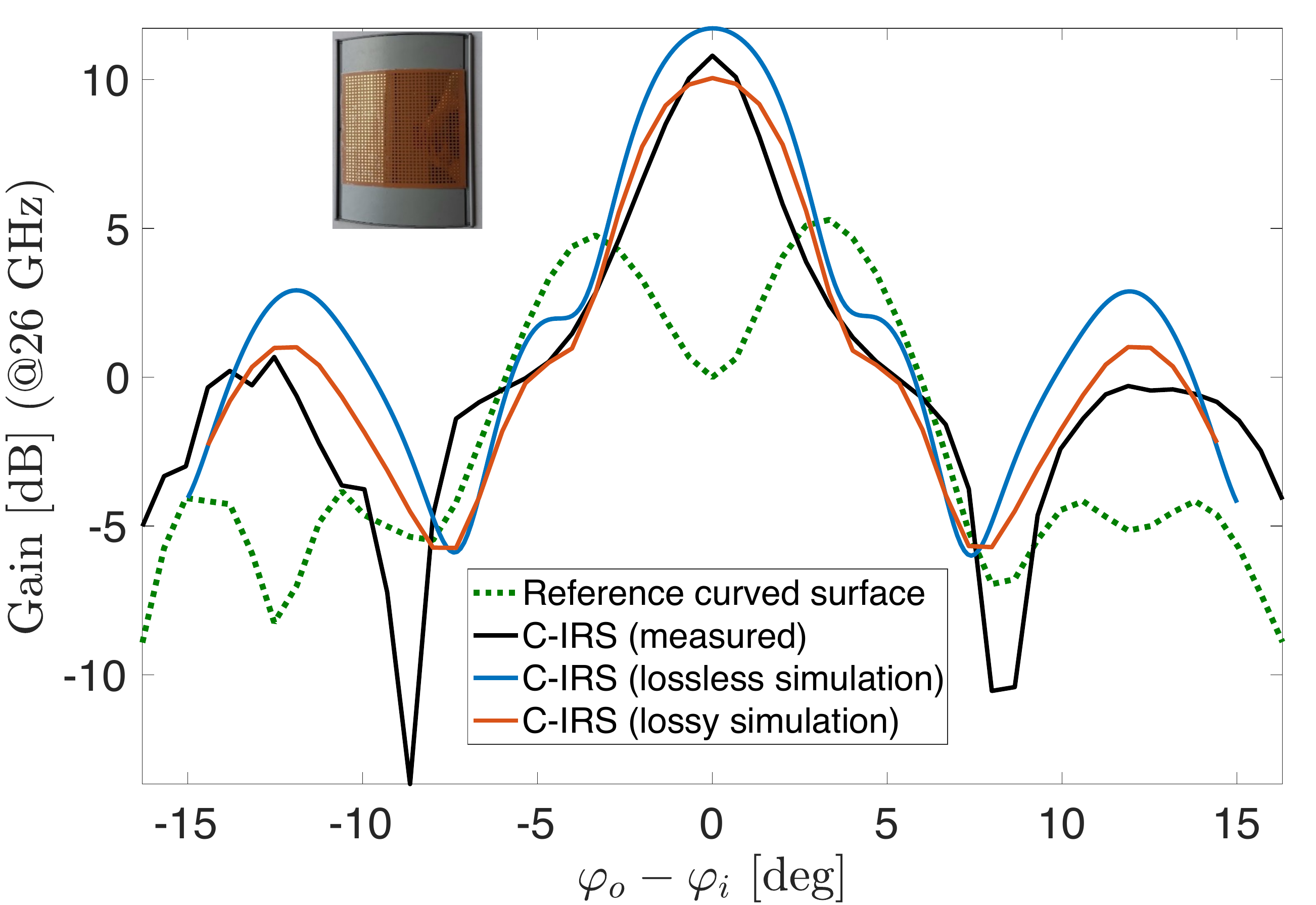}}
	\caption{The measured scattered field of the C-IRS compared with the reference curved surface. Design frequency is 26 GHz.}\label{fig:G2}
\end{figure}

The realized structure has been measured in anechoic chamber. A transmitting antenna is placed at 1.5 m from the surface and the receiving antenna at the same distance is moved horizontally to collect the scattered field over a 1 m extension, corresponding to about $\pm 18 $ deg from broadside. An accurate positioning of the reference target and the patterned target allowed to compare the two scattered field in the same conditions, so as to verify the focusing capabilities. For clarity, a picture of the measurement setup is shown in Fig. \ref{fig:G3} and a detail of the C-IRS mounted in the chamber is in Fig. \ref{fig:G3a}. It must be noted that we are still in Fresnel region of the scattered field, but even so, the widely different behavior of the simple curved surface with respect to the patterned one is clearly visible in the results in Fig. \ref{fig:G2}.
This compares the scattered field of the plain curved surface (labelled reference curved surface) with the one of the C-IRS as a function of the position of the receiving antenna, that is represented by angle $\varphi_o$. The incidence angle is  $\varphi_i$.
The gain obtained by the C-IRS in terms of focusing of the scattered field is clearly visible and it amounts to about 10 dB. In the comparison, a very good agreement between simulated data and measured data can be observed. Thanks to the robustness in the design, tolerances and possible difference of the material permittivity with respect to nominal value ($\varepsilon_r = 4.6$) play a minor role. 
The comparison in Fig. \ref{fig:G2} is also useful to assess the performances in terms of ohmic efficiency. EM simulations with and without losses indicate that the impact of losses  is relatively small and the advantage of patterned C-IRS is still very clear in the figures.

\section{Simulation Results in V2V Networks}\label{sect:V2X_results}

The goal is now to evaluate the advantages of C-IRS lodged on CAV doors in a multi-lane scenario of $500$ m length with $N_L=5$ lanes, each of $W_L = 5$ m width. Numerical analysis is y evaluating the average SNR. All vehicles in the scenario are modelled with an occupation region of $5 \times 1.8 \times 1.5 $ m$^3$ and randomly distributed according to a point Poisson process with parameter (i.e., traffic density) $\rho$ cars/km per lane. Simulation parameters are in Table \ref{tab:SimParam}. We consider a variable percentage of CAVs $P$ in the scenario, as expected within the transition towards autonomous driving. While CAVs are equipped with C-IRS, other vehicles are not.

Let us assume that Tx and Rx CAVs know their positions, namely $\mathbf{p}_{tx}$ and $\mathbf{p}_{rx}$, as well as the position of other CAVs in the scenario. In the present simulations, we assume for simplicity Tx and Rx located on the same lane at distance $r_d=100$ m. When C-IRS are designed for specular reflection, a candidate relaying CAV can only be located halfway between Tx and Rx, in $\mathbf{p}_{rel} \in \mathcal{A}_{rel}$, where $\mathcal{A}_{rel}$ is set of allowed relay positions, i.e., a spatial region of size $A_{rel} = W_{rel} \times L_{rel}$ centered in $(\mathbf{p}_{rx} - \mathbf{p}_{rx})/2$. While $W_{rel} = N_\ell W_\ell$ is the highway width, $L_{rel}$ is chosen to be two times the length of the C-IRS, i.e., $L_{rel} = 2L$. The best beam pair between Tx and Rx $(\mathbf{f},\mathbf{w})$, to be used in \eqref{eq:receivedSignal}, is chosen within dynamic codebooks $\mathcal{F}$ and $\mathcal{W}$, built upon the knowledge of the $C$ candidate relaying CAVs within $\mathcal{A}_{rel}$. Thus, 
\begin{equation}\label{eq:optimalBeam}
    (\mathbf{f}_{opt}, \mathbf{w}_{opt}) = \underset{\substack{\mathbf{f}\in\mathcal{F}\\\mathbf{w}\in\mathcal{W}}}{\mathrm{argmax}}\,\left\{ \frac{\sigma^2_s\lvert\mathbf{w}^\mathrm{H}\mathbf{H}\mathbf{f}\rvert^2}{K\sigma^2_n}\right\}.
\end{equation}
where $\lvert\mathcal{F}\rvert = \lvert\mathcal{W}\rvert = C +1 $, i.e., the cardinality is the number of candidate relays plus the direct link. The choice in \eqref{eq:optimalBeam} is therefore dictated by the maximum SNR at the Rx via all the possible links (direct and relayed). Notice that $C$ is ruled by both the traffic density $\rho$ and the CAV percentage $P$. Fixing $C$ and the geometry of the scenario, a blockage event on a single Tx-relay-Rx link happens when at least one of the two Tx-relay and relay-Rx links is blocked. The blockage probability is therefore the probability that all the Tx-relay-Rx links are jointly blocked. The presence of one or more vehicles interposing between Tx-Rx, Tx-relay and relay-Rx is accounted as additional power loss on the link budget compared to the usual free-space loss, as indicated in \cite{dong2021vehicular}.

\begin{figure}
    \centering
    \subfloat[][$\rho=10$ cars/km]{\includegraphics[width=0.9\columnwidth]{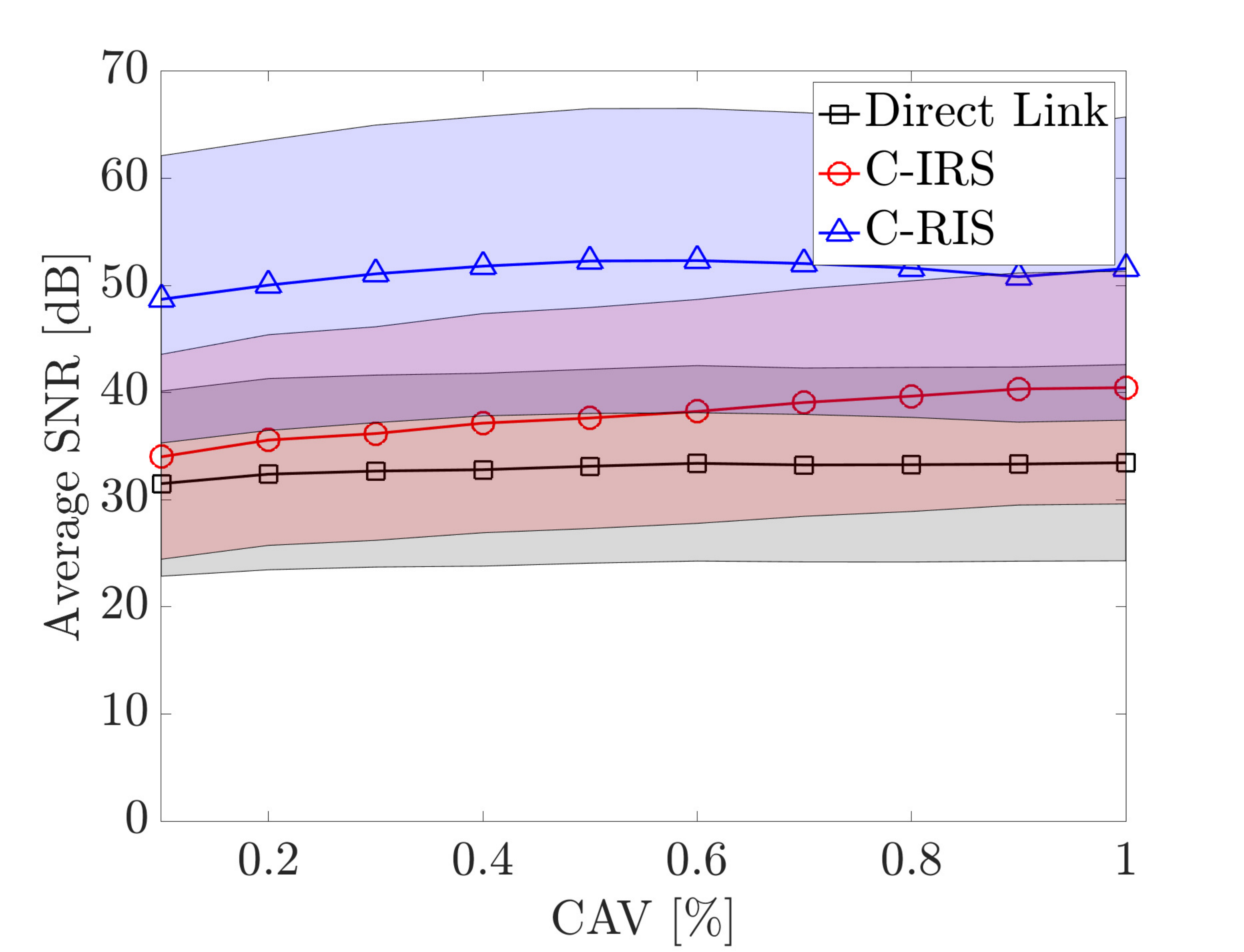}\label{subfig:SNR_rho10}}\vspace{-0.25cm}\\
    \subfloat[][$\rho=50$ cars/km]{\includegraphics[width=0.9\columnwidth]{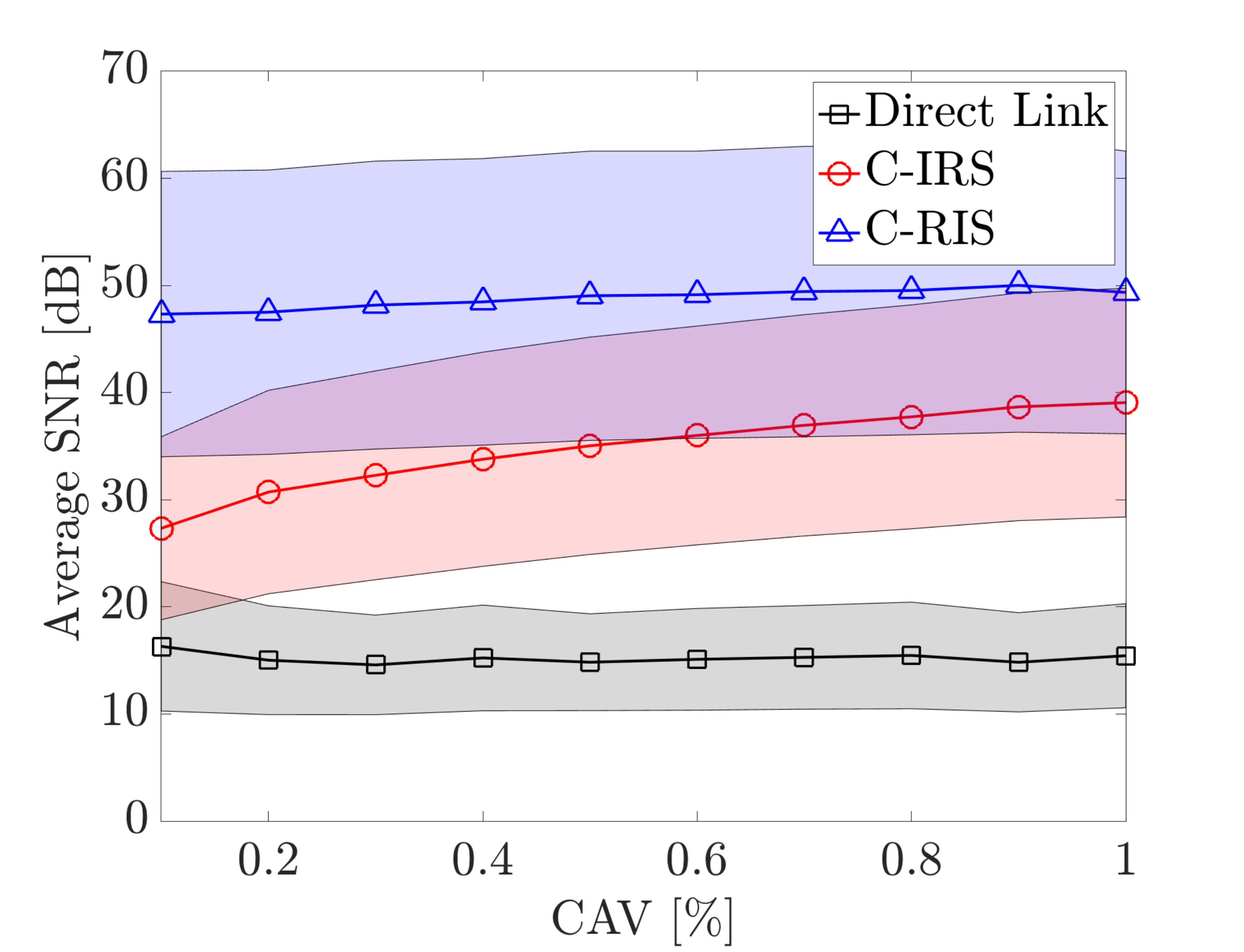}\label{subfig:SNR_rho50}}
    \caption{Average SNR varying the CAV percentage $P$ in the scenario, for (\ref{subfig:SNR_rho10}) $\rho=10$ cars/km (\ref{subfig:SNR_rho50}) $\rho=50$ cars/km }
    \label{fig:SNR}
\end{figure}

Figure \ref{fig:SNR} shows the trend of the average SNR varying the CAVs percentage $P$, for two different traffic densities, $\rho=10$ cars/km (Fig. \ref{subfig:SNR_rho10}, low-traffic) and $\rho=50$ cars/km (Fig. \ref{subfig:SNR_rho50}, medium-to-high traffic). The shaded areas represent the confidence interval defined by the standard deviation. We compare the performance of a V2V system relying only on the direct link (black curves) to the proposed C-IRS-assisted one (red curves). To mere comparison, we also report the case in which CAVs are equipped with C-RIS (blue curves) \cite{mizmizi2022conformal}, i.e., they can dynamically configure the RIS phase to enable arbitrary reflections by assuming the perfect knowledge of CAVs positions. In this latter case, the candidate relays can be searched within \textit{all} the scenario, thus providing the upper performance bound. Compared to a conventional V2V link, the C-IRS-assisted one provides a remarkable SNR gain that increases with $P$ and $\rho$. In particular, for $\rho=10$ cars/km the SNR gain is $\approx 8$ dB, while for $\rho=10$ cars/km it amounts to $\approx 25$ dB, for $P\rightarrow 100\%$. For increasing $\rho$, moreover, C-IRS almost fill the gap provided by C-RIS: this is a direct consequence of the increase in potential relaying CAVs equipped with C-IRS. For C-RIS, instead, it is sufficient to find one relay within the scenario to enable a strong reflection link (the RIS gain is approx $50$ dB, at least in favourable conditions, i.e., perfect phase configuration), thus the performance are substantially independent from $\rho$ and $P$. These results justify the research effort in vehicle-lodged metasurfaces for 6G V2V networks, and particularly toward inexpensive and pre-configured C-IRS, that do not need dedicated control signaling and power supply.

\begin{table}[b!]
    \centering
    \caption{Simulation Parameters}
    \begin{tabular}{l|c|c}
    \toprule
        \textbf{Parameter} &  \textbf{Symbol} & \textbf{Value(s)}\\
        \hline
        Carrier frequency & $f$  & $26$ GHz \\
        Number of Tx/Rx antennas & $K$ & $8$\\
        Tx-Rx distance & $r_d$ & $100$ m\\
        Number of C-IRS elements & $M \times N$ & $400 \times 400$ \\
        C-IRS element spacing & $d_m, d_n$ & $\lambda/4$\\
        C-IRS curvature radius & $R$ & $8$ m\\
        C-IRS configuration param. &$\overline{\theta}$ & $80$ deg\\
        Transmitted power & $\sigma^2_s$ & $20$ dBm\\
        Noise power & $\sigma^2_n$ & $-88$ dBm\\
        \bottomrule
    \end{tabular}
    \label{tab:SimParam}
\end{table}

\section{Conclusion}\label{sect:conclusion}

This paper proposes the usage of C-IRS, properly lodged on CAVs' sides, to mitigate the blockage affection in V2V networks without involving any active relaying. We analytically provide the C-IRS phase configuration under a cylindrical approximation of cars' doors, and validate the derived models by physical realization and testing in anechoic chamber of a C-IRS operating at 26 GHz frequency. 
The measured reflection pattern along the the curved plane of the C-IRS shows a perfect match with the predicted one by both lossless and lossy EM simulations. In addition, numerical results in a multi-lane highway V2V scenario show the SNR gain when equipping CAVs with C-IRS, increasing with both CAV's percentage and traffic density, attaining a remarkable 20 dB compared to the usage of direct V2V link only.

\section*{Acknowledgment}
This work has been carried out within the Huawei-Politecnico di Milano Joint Research Lab.

\bibliographystyle{IEEEtran}
\bibliography{biblio}

\end{document}